\documentclass[fleqn,usenatbib]{mnras_tex_edited}

\usepackage{amssymb,amsmath,verbatim,mathtools,needspace,enumitem,etoolbox,graphicx,physics,microtype,natbib,url,hyperref,mathrsfs,bm,ulem,orcidlink}

\renewcommand{\emph}[1]{\textit{#1}}
\newcommand{\approptoinn}[2]{\mathrel{\vcenter{
  \offinterlineskip\halign{\hfil$##$\cr
    #1\propto\cr\noalign{\kern2pt}#1\sim\cr\noalign{\kern-2pt}}}}}

\DeclareRobustCommand{\okina}{%
  \raisebox{\dimexpr\fontcharht\font`A-\height}{%
    \scalebox{0.8}{`}%
  }%
}

\newcommand{\msun}{{\rm M}_\odot}
\newcommand{\vtwo}[1]{\textcolor{black}{#1}} 
\newcommand{\vthree}[1]{\textcolor{black}{#1}}

\title[Can GW231123 have a stellar origin?]{\vtwo{Can GW231123 have a stellar origin?}}

\author[Croon et al.]{%
Djuna Croon$\,$\orcidlink{0000-0003-3359-3706}$^{1}$\thanks{Authors are listed alphabetically.},
Davide Gerosa$\,$\orcidlink{0000-0002-0933-3579}$^{2,3}$, 
Jeremy Sakstein$\,$\orcidlink{0000-0002-9780-0922}$^{4}$
\medskip
\\
$^{1}$ Institute for Particle Physics Phenomenology, Department of Physics, Durham University, Durham DH1 3LE, UK\\
$^{2}$Dipartimento di Fisica ``G. Occhialini'', Universit\'a degli Studi di Milano-Bicocca, Piazza della Scienza 3, 20126 Milano, Italy\\
$^{3}$INFN, Sezione di Milano-Bicocca, Piazza della Scienza 3, 20126 Milano, Italy\\
$^{4}$Department of Physics \& Astronomy, University of Hawai\okina i, Watanabe Hall, 2505 Correa Road, Honolulu, HI, 96822, USA
}

\pubyear{\the\year{}}

\begin{document}
\label{firstpage}
\pagerange{\pageref{firstpage}--\pageref{lastpage}}

\maketitle

\begin{abstract}

The gravitational wave event GW231123 detected by the LIGO interferometers during their fourth observing run features two black holes with source-frame masses of \vtwo{$137^{+23}_{-18}\msun$ and $101^{+22}_{-50}\msun$} --- \vtwo{in the range of the pair-instability black hole mass gap predicted by standard stellar evolution theory}.~Both black holes are also inferred to be rapidly spinning ($\chi_1 \simeq 0.9$, $\chi_2 \simeq 0.8$).~The primary object in GW231123 is the heaviest stellar mass black hole detected to date, which, together with its extreme rotation, raises questions about its astrophysical origin.
Accounting for the unusually large spin of $\sim 0.9$ with hierarchical mergers requires some degree of fine tuning.
We investigate whether such a massive, highly spinning object could plausibly form from the collapse of a single rotating massive star.~We simulate stars with an initial core mass of $160\,\msun$~---~sufficient to produce BH masses at the upper edge of the 90\% credible interval for $m_1$ in GW231123 --- across a range of rotation rates and $^{12}\mathrm{C}(\alpha,\gamma)^{16}\mathrm{O}$ reaction rates.
~We allow for differential rotation to explore the high-spin regime.
~\vtwo{In this limit of weak angular momentum transport,} we find that:~(i) rotation shifts the pair-instability mass gap to higher masses, introducing an important correlation between masses and spins in gravitational wave predictions;~and (ii) highly spinning BHs with masses $\gtrsim 150 \rm M_\odot$ can form above the mass gap.~Our results suggest that the primary object of GW231123 may be the first directly observed black hole that formed via direct core collapse following the photodisintegration instability.~

\end{abstract}

\begin{keywords}
black hole mergers --- gravitational waves
\end{keywords}

\section{Introduction}
On 23 November 2023, the LIGO Hanford and Livingston interferometers detected a short, $\sim5$‐cycle signal consistent with the merger of two black holes (BHs) with source‐frame masses 
\begin{equation}
   \vtwo{ m_1 = 137^{+23}_{-18}\,\msun,\quad
m_2 = 101^{+22}_{-50}\,\msun,}
\end{equation}
at redshift $z\approx0.39$~\citep{LIGOScientific:2025rsn}.~This event, dubbed GW231123, has total mass ($190$–$265\,\msun$) and spins ($\chi_1\simeq0.90^{+0.10}_{-0.19}$, $\chi_2\simeq0.80^{+0.20}_{-0.51}$), pushing %
 into a regime where waveform systematics still need to be fully understood~\citep{LIGOScientific:2025rsn}.~Yet its high signal‐to‐noise ratio ($\sim22.5$) %
and two‐detector coincidence make its identification robust.~The high masses and spins distinguish GW231123 as a qualitatively new class of system, raising questions about its possible (astro)physical origin;~see e.g., \citet{Stegmann:2025cja,Yuan:2025avq,Li:2025fnf,Cuceu:2025fzi,Tanikawa:2025fxw,Bartos:2025pkv,Kiroglu:2025vqy,Paiella:2025qld}.

\vtwo{
Ostensibly, the combination of large masses and large spins in GW231123 could be explained by binaries formed through hierarchical mergers (for context, see \citealt{Gerosa:2021mno} and references therein) and this appears to be a leading explanation in some of the recent litterature~\citep{LIGOScientific:2025rsn,Stegmann:2025cja,Li:2025fnf,Paiella:2025qld}.~We argue that a hierarchical merger origin for GW23112's primary BH is unlikely; cf. \citet{Bartos:2025pkv} for a similar argument.~Hierarchical mergers in dense stellar environments produce a spin distribution that is sharply peaked at $\chi \sim 0.7$~\citep{Gerosa:2017kvu,Fishbach:2017dwv}, which matches the primary BH spin in GW231123 only near the lower edge of its 90\% credible interval.~Spins of $\chi \sim 0.9$ are too large to be explained as merger remnants~\citep{Gerosa:2021hsc}, requiring substantial fine‐tuning of the progenitor binary (involving either a somewhat extreme mass ratio and/or mostly aligned spins).~Motivated by these considerations, in this work we explore the possibility that the primary component of GW231123 could have formed from the direct collapse of a rotating massive star.}

Stellar evolution theory  predicts a ``pair‐instability" black hole mass gap (BHMG) between approximately $60$ and $130\,\msun$, driven by electron-positron pair production in helium cores that powers (pulsational) pair‐-instability supernovae (hereafter (P)PISN) and completely disrupts stars with core masses up to $\sim135\msun$~\citep{Fowler:1964zz,Barkat:1967zz,1967ApJ...148..803R,1968Ap&SS...2...96F,Spera:2017fyx,Woosley:2002zz,Woosley:2016hmi,2019ApJ...887...53F,Farmer:2020xne,Mehta:2021fgz,Farag:2022jcc, Woosley:2021xba, Hendriks:2023yrw}.~Various stellar parameters and other effects determine the location of the BHMG, most importantly the $^{12}\mathrm{C}(\alpha,\gamma)^{16}\mathrm{O}$ rate~\citep{Takahashi:2018kkb,2019ApJ...887...53F,Sakstein:2020axg,Farmer:2020xne,Mehta:2021fgz,Farag:2022jcc}, which can give variations of $\sim47\msun$~\citep{Mehta:2021fgz} in the location of the upper and lower edges of the BHMG.~

If it had zero spin, the primary BH in GW231123 would likely lie above the BHMG unless the $^{12}\mathrm{C}(\alpha,\gamma)^{16}\mathrm{O}$ reaction rate were $\sim3\sigma$ smaller than its median value~\citep{Mehta:2021fgz}.~However, this is based on simulations of (P)PISN which have neglected rotation, so this conclusion is unlikely to apply to GW231123 and its extremely high spins.~Indeed, previous theoretical work~\citep{Marchant:2020haw} has shown that rotation can raise the \emph{lower} edge of the gap by $\sim 15\%$ (from $45.5\msun$ to $52.4\msun$).~It is {currently} not known how rotation affects the \emph{upper} edge of the BHMG.

\vtwo{A major uncertainty in modeling the rotational evolution of very massive stars is the efficiency of internal angular momentum transport. Much of the empirical support for strong core-envelope coupling, often modeled via the Spruit-Tayler (ST) dynamo, comes from low- and intermediate-mass stars. Its operation in very massive, radiation-dominated, and rapidly evolving stars, however, remains largely unconstrained observationally. The high spin inferred for the primary of GW231123 would be difficult to reconcile with standard stellar evolution models that enforce near-rigid rotation throughout the star’s lifetime. Motivated by this tension, we explore an optimistic bracketing scenario in which internal angular momentum transport is inefficient, allowing the core to retain substantial rotation up to collapse. This assumption is not meant to be representative of typical massive stars, but rather to test whether the photodisintegration formation channel can remain viable even in the extreme case required by GW231123. Finally, we note that several studies have proposed evolutionary scenarios in which progenitors retain or regain substantial core rotation, including chemically homogeneous evolution \citep{deMink:2016vkw,Marchant:2016wow,Popa:2025dpz}, early envelope stripping in binaries \citep{Qin:2018vaa}, and tidal spin-up \citep{Olejak:2021iux}. Our results provide a first indication of how pair-instability boundaries might be modified in such scenarios, independent of the detailed mechanism by which the core is spun up.}

\section{Core collapse instabilities}
The predicted BHMG is a direct result of the physics of pair‐instability, a well-studied effect ~\citep{Fowler:1964zz,Barkat:1967zz,1967ApJ...148..803R,1968Ap&SS...2...96F,Spera:2017fyx,Woosley:2002zz,Woosley:2016hmi,Takahashi:2018kkb,2019ApJ...887...53F,Farmer:2020xne,Mehta:2021fgz,Farag:2022jcc, Woosley:2021xba, Hendriks:2023yrw} in massive post-main-sequence stars.~At core temperatures $\gtrsim 10^{9}\,\mathrm{K}$, high‐energy photons convert into non-relativistic electron-positron pairs, \vtwo{softening the equation of state such that the adiabatic index $\Gamma_1<4/3$, causing the core to contract, which in turn leads to explosive nuclear burning, primarily of oxygen.}~
If the released nuclear energy exceeds the gravitational binding energy, the star is completely disrupted with no BH remnant.~In a slightly lower mass range the explosion is weaker and instead ejects the outer layers in pulses (a PPISN), removing mass until the remaining core collapses to a lighter BH.
Above a critical helium core mass ($\sim 130 \msun$), stellar cores become so hot {($T_c \gtrsim 9 \times 10^9\,\mathrm{K}$)}
that high‐energy photons begin to photodisintegrate iron‐group nuclei into $\alpha$‐particles and free nucleons.~This endothermic process decreases pressure support in the core \vtwo{(relative to continued nuclear burning)}, accelerating collapse --- a phenomenon known as the photodisintegration instability.~In this scenario, the star implodes directly to a BH, defining the upper edge of the BHMG.~

\vtwo{~Rotation impacts pair instability through two related mechanisms.~First,}
the extra support provided by the centrifugal force results in lower temperatures \vtwo{at a given central density} compared with non-rotating stars~\citep{Marchant:2020haw,Huynh:2025fnt}.
\vtwo{This shifts the star away from the regime where the effective adiabatic index drops below the critical value of $4/3$ for dynamical instability.~Second, if the core contracts while approximately conserving angular momentum, it spins up, further increasing centrifugal support during contraction, effectively lowering the critical value of $\Gamma_1$ for stability below $4/3$.~Together, these effects act to stabilize rapidly rotating stars against pair instability.~It is presently not known what the effect on photodisintegration instability is.}

{In the absence of rotation,} the onset and outcome of (P)PISN depend sensitively on the rates of core helium‐burning reactions, particularly the competition between the triple‐$\alpha$ process, which produces $^{12}$C,  and the subsequent $^{12}\mathrm{C}(\alpha,\gamma)^{16}\mathrm{O}$ channel.~While uncertainties in both reaction rates affect the carbon-to-oxygen ratio, variations in the $^{12}\mathrm{C}(\alpha,\gamma)^{16}\mathrm{O}$ rate are the most important, dominating the shift in the carbon-oxygen core mass at which electron-positron pair production induces dynamical instability and thus altering the mass thresholds for (P)PISN events~\citep{Takahashi:2018kkb,2019ApJ...887...53F,Farmer:2020xne,Mehta:2021fgz,Farag:2022jcc}.~This rate is highly uncertain, with current experiments disagreeing by statistically significant amounts~\citep{deBoer:2017ldl}.~It is therefore standard practice to consider $\pm3\sigma$ variations about the median between the various experimental constraints~\citep{2016MNRAS.456.3866C,deBoer:2017ldl,Takahashi:2018kkb,2019ApJ...887...53F,Farmer:2020xne,Mehta:2021fgz,Farag:2022jcc,2022ApJ...935...21C,2023ApJ...954...51C,Croon:2023kct}.~ 

As the $^{12}\mathrm{C}(\alpha,\gamma)^{16}\mathrm{O}$ rate is raised, a larger fraction of $^{12}$C is converted into $^{16}$O during core helium burning.~{This higher oxygen abundance results in a stronger explosion, whilst the lower $^{12}$C abundance implies less carbon is available to form a convective $^{12}$C burning shell to counteract contraction~\citep{Farmer:2020xne}.}
Consequently, both the \emph{lower} and \emph{upper} edges of the mass gap move downward in tandem, preserving a nearly constant gap width of~$\Delta M_{\rm BHMG}\approx80^{+9}_{-5}\msun$~\citep{Mehta:2021fgz}.~Across the full $\pm 3 \sigma$ uncertainty in the reaction rate, the lower boundary shifts from $\sim59^{+34}_{-13}\msun$ down to lighter values, while the upper boundary moves from $\sim139^{+30}_{-14}\msun$ to correspondingly lower masses~\citep{Mehta:2021fgz}.~

\section{Simulation suite} 
To assess the impact of rotation and nuclear physics on the boundaries of the pair-instability regime, we computed a grid of stellar models with \vtwo{an initial helium core mass of $160 \msun$}.~This value was chosen such that, if no PISN occurs, the star can be expected to form a BH with mass at the edge of the measured $90\%$ confidence for $m_1$ in GW231123.~We vary the stellar rotation rate $\Omega/\Omega_{\mathrm{crit}} $ between $0$ and $1$ to capture the effects of increasing centrifugal support.~The critical angular velocity 
\begin{equation}
\Omega_{\rm crit}
\equiv \sqrt{\frac{(1 - \Gamma)\,G\,M}{R_{\rm eq}^3}}\,,
\end{equation}
 is defined as the rotation rate at which the outward centrifugal acceleration at the stellar equator exactly balances the inward effective gravitational acceleration, where $M$ is the gravitational mass, $R_{\rm eq}$ the equatorial radius, and $\Gamma \equiv L/L_{\rm Edd} = L\kappa /{4\pi G M c}$ is the local Eddington factor which accounts for radiation pressure.~Additionally, we explore the variation of the rate of the {temperature-dependent} $^{12}\mathrm{C}(\alpha,\gamma)^{16}\mathrm{O}$ reaction over $\pm3\sigma$ {from its median rate $R_{\rm med}(T)$.~Here, $\sigma$ parametrizes the variation as $R_\sigma(T)=R_{\rm med}(T)\exp[\sigma\mu(T)]$ with $\mu(T)$ the uncertainty at each temperature point, which is assumed to follow a
log-normal distribution~\citep{deBoer:2017ldl,Farmer:2020xne,Mehta:2021fgz}.}

Our simulations were performed using the stellar structure code MESA, version 15140~\citep{Paxton:2010ji,Paxton:2013pj,Paxton:2015jva,Paxton:2017eie,Paxton:2019lxx,MESA:2022zpy}.~While MESA is a one-dimensional code --- meaning that it assumes spherical symmetry --- it is capable of simulating rotating objects via the shellular approximation, where the radial coordinate $r$ is replaced with the volume-equivalent radius of an isobar.~A comprehensive description of rotation in MESA is provided in \citet{Paxton:2019lxx}.~Our code can be found in the paper's reproduction package \citep{sakstein_2025_16898502} at the following URL:~\href{https://zenodo.org/records/16898502}{https://zenodo.org/records/16898502}. 

\vtwo{We simulate $160\msun$ rotating helium cores with metallicity $Z=10^{-5}$ from the zero-age horizontal branch (ZAHB) to their end-state --- either core collapse or disruption due to PISN --- since their hydrogen envelopes are expected to be stripped by strong stellar winds or binary interactions, leaving behind bare helium cores that determine the subsequent evolution~\citep{Heger:2001cd,Eldridge:2007mi,Belczynski:2016obo,Woosley:2016hmi}.~Rotation is initialized at the ZAHB.~Mass loss to Wolf-Rayet winds are described following the prescription of \citet{Brott:2011ni}, with a rotational enhancement described in \citet{Paxton:2013pj}.~Stars that approach critical rotation are subject to additional mass losses following \citet{Paxton:2013pj} that remove angular momentum to keep them sub-critical.~Convective mixing is implemented following the time-dependent formulation of \citet{Marchant:2018kun,Renzo:2020rzx}, which captures rapid changes in convective criteria during phases of dynamical instability but reduces to standard mixing length theory on long time-scales.~We take the mixing length efficiency to be $\alpha_{\rm MLT}=2.0$.~Convection is determined using the Ledoux criterion with semi-convection efficiency $\alpha_{\rm SC} = 1.0$.~We employ exponential overshooting with $f=0.01$.~The switch from convective mixing to overshooting occurs at a distance $f_p H_P$ into the convective zone with $f_p=0.005$ and $H_p$ the pressure scale-height.~We adopted the AGSS09 metal fractions \citep{2009ARA&A..47..481A}.~These choices follow standard practice in recent studies of massive helium star evolution and pair-instability supernovae \citep{Croon:2020oga,Sakstein:2020axg,Straight:2020zke,Croon:2020ehi,Sakstein:2022tby,Croon:2023trk}, including the rotation prescription introduced by \citet{Marchant:2020haw}.~We replaced the default MESA  $^{12}\mathrm{C}(\alpha,\gamma)^{16}\mathrm{O}$ rate with the state-of-the-art tabulated values from \citet{Mehta:2021fgz}, which sample the \citet{deBoer:2017ldl} rates;~again consistent with contemporary works~\citep{Mehta:2021fgz,Farag:2022jcc,Croon:2023kct}.~Our resolution is higher than most MESA PPISN studies, following ecent works that suggest such high resolutions are needed to fully resolve the core collapse-PPISN transition~\citep{Mehta:2021fgz,Farag:2022jcc,Croon:2023kct}, although we note that a resolution study on the PISN-photodisintegration boundary has yet to be performed.~Specifically, we set {\tt delta\_lgRho\_cntr\_limit = 0.001d0} and {\tt max\_dq = 5d-4 } to enforce a smaller time-step \vthree{and higher spatial resolution.}}

\vtwo{
As noted above, our study is motivated by proposed evolutionary scenarios in which black hole progenitor stars are rapidly rotating. For this reason, we do not include the ST dynamo \citep{Spruit:2001tz}, which is commonly invoked in stellar models to enforce near-rigid rotation through efficient angular momentum coupling between stellar layers \citep{Fuller:2019sxi,Fuller:2022ysb}. The physical validity and efficiency of this mechanism remain uncertain, and several studies have questioned whether the magnetic instabilities required to sustain the ST dynamo can operate under realistic stellar conditions \citep[e.g.][]{Denissenkov:2006tk,Zahn:2007uk}.
}

\vtwo{
At the same time, it is widely recognized that some form of efficient angular momentum coupling must operate in stars in order to reproduce observational constraints, particularly those inferred from low- and intermediate-mass stellar remnants, such as the slow rotation of white dwarfs and neutron stars and the modest core--envelope differential rotation observed in evolved stars. Whether similarly efficient coupling operates in the very massive stars considered here, especially in the pair-instability regime, remains uncertain.
}

\vtwo{Models that adopt efficient angular momentum transport via the ST dynamo predict nearly non-rotating black holes \citep{Marchant:2020haw}, which are difficult to reconcile with the high component spins inferred for GW231123, potentially pointing to weaker internal angular momentum transport in its progenitor.\footnote{We note that alternative interpretations suggest the nominal spin inference itself may be biased \citep{Ray:2025rtt}.} We therefore allow for differential rotation and spin retention in our models, which both enables high-spin remnants and directly impacts the onset of pair-instability by shifting the upper edge of the mass gap to higher masses. Demonstrating that a rapidly spinning primary can still form above the mass gap strengthens the plausibility of a stellar-origin scenario involving photodisintegration collapse.}

\begin{figure}
    \centering
    \includegraphics[width=\linewidth]{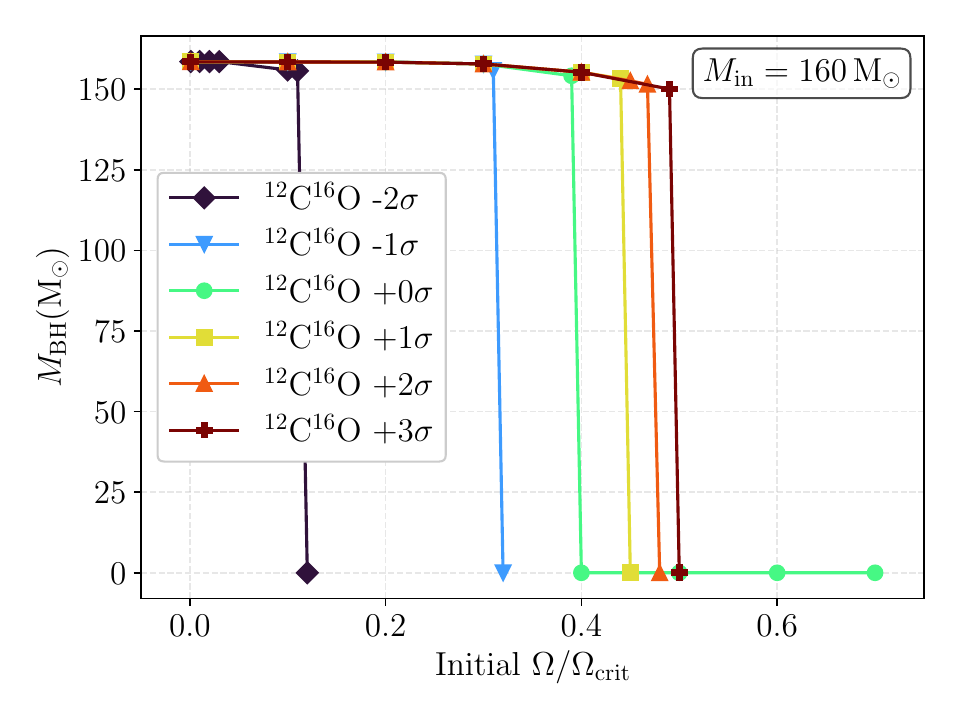}
    \includegraphics[width=\linewidth]{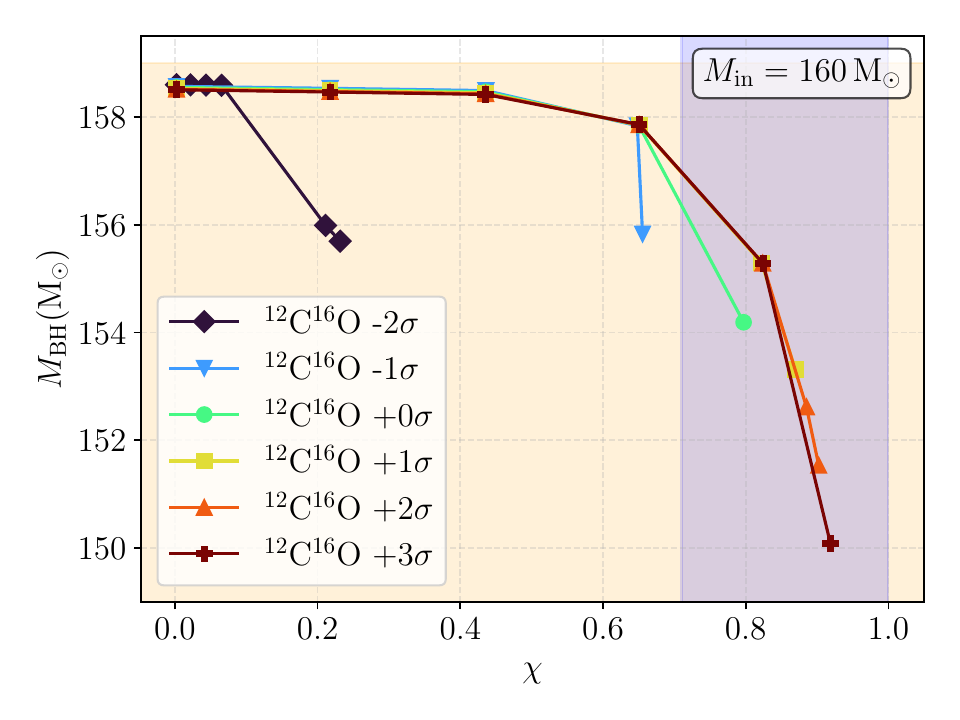}
    \caption{Final BH mass as a function of initial rotation with initial helium core mass $160 \rm \msun$, for several values of the $^{12}\mathrm{C}(\alpha,\gamma)^{16}\mathrm{O}$ rate.~The top panel shows results as a function of the initial stellar rotation $\Omega$ in units of the critical value $\Omega_{\rm crit}$;~the bottom panel shows results as a function of the BH spin $\chi$.~Stars with faster rotation and smaller  $^{12}\mathrm{C}(\alpha,\gamma)^{16}\mathrm{O}$ reaction rate undergo PISN and are not shown in the figure.~The shaded contours indicate the $90\%$ credible intervals of the primary mass (orange) and spin (purple) in GW231123 \protect\citep{LIGOScientific:2025rsn}.~%
    }
    \label{fig:BHMassSpin}
\end{figure}

Some simulations feature spin parameters exceeding unity at core-collapse.~Although the BH may form with maximal spin, further accretion can reduce its spin if the added mass is not accompanied by sufficient angular momentum, or if excess angular momentum is expelled by magnetohydrodynamic feedback \citep[e.g.][]{McKinney:2012vh}).~To estimate the final spin of the BH,  we adopt a version of the model developed by~\cite{Batta:2019clm}, consistent with previous work~\citep{Marchant:2020haw}.~We assume that the innermost $3\msun$ of the progenitor collapses promptly into a BH with maximal spin.~{We have explicitly verified that the results are insensitive to the size of this core, see \vtwo{appendix \ref{sensitivitytocollapse}} for more details.~The remainder of the star is assumed to either fall directly into the BH or accrete through a disk, delivering additional mass and angular momentum.~Accretion is regulated by feedback, ensuring that the final spin does not exceed the Kerr limit ($\chi \leq 1$).~

\section{Results}
\subsection{Impact of rotation on the black hole mass gap}
We show the results of our simulations in Fig.~\ref{fig:BHMassSpin}.~The top panel shows the final BH mass as a function of the initial rotation rate $\Omega/\Omega_{\rm crit}$ for a $160\,M_\odot$ helium core, with ${}^{12}\mathrm{C}(\alpha,\gamma){}^{16}\mathrm{O}$ reaction rates varied from $-2\sigma$ to $+3\sigma$.~The $-3\sigma$ simulations produced PISN with no BH remnants for all $\Omega/\Omega_{\rm crit}$ (including zero) and are thus not shown.~At low $\Omega/\Omega_{\rm crit}$ all models undergo direct collapse, yielding $M_{\rm BH}\approx159\,M_\odot$.~At some critical rotation threshold $\Omega_{\rm PISN}$, the pair‐instability disrupts the star completely and no remnant is left.~This threshold increases from $\Omega_{\rm PISN}\sim0.1 \Omega_{\rm crit}$ for the $-2\sigma$ rate to $\Omega_{\rm PISN}\sim0.5 \Omega_{\rm crit}$ for the $+3\sigma$ rate, demonstrating that stronger carbon burning stabilizes the core against pair‐instability to higher rotation.

The bottom panel of Fig.~\ref{fig:BHMassSpin} presents these same models in the BH spin-mass plane.~Stars with a helium core mass of $160 \msun$ cannot form BHs with a spin above a certain threshold, which depends on the ${}^{12}\mathrm{C}(\alpha,\gamma){}^{16}\mathrm{O}$ reaction rate.~In particular, BHs with spins and masses compatible with the {$90\%$ confidence interval of the} primary in GW231123 %
cannot be formed if the ${}^{12}\mathrm{C}(\alpha,\gamma){}^{16}\mathrm{O}$ reaction rate is smaller than its measured median value.

\vtwo{As anticipated above, our results demonstrate that the photodisintegration boundary which marks the upper edge of the BHMG shifts to higher stellar masses in the presence of rapid rotation.}~As shown in Fig.~\ref{fig:photodisintegration}, rotating stars reach lower core temperatures $T_c$ before the onset of explosive oxygen burning leads to rapid expansion and a temperature drop.~Thus, these stars
avoid reaching the conditions required for photodisintegration-induced collapse, and instead undergo PISN.~For the median $^{12}\mathrm{C}(\alpha,\gamma)^{16}\mathrm{O}$ rate, we find that this occurs at $\Omega_{\rm PISN} \sim 0.4 \Omega_{\rm crit}$.

\begin{figure}
\includegraphics[width=0.95\linewidth]{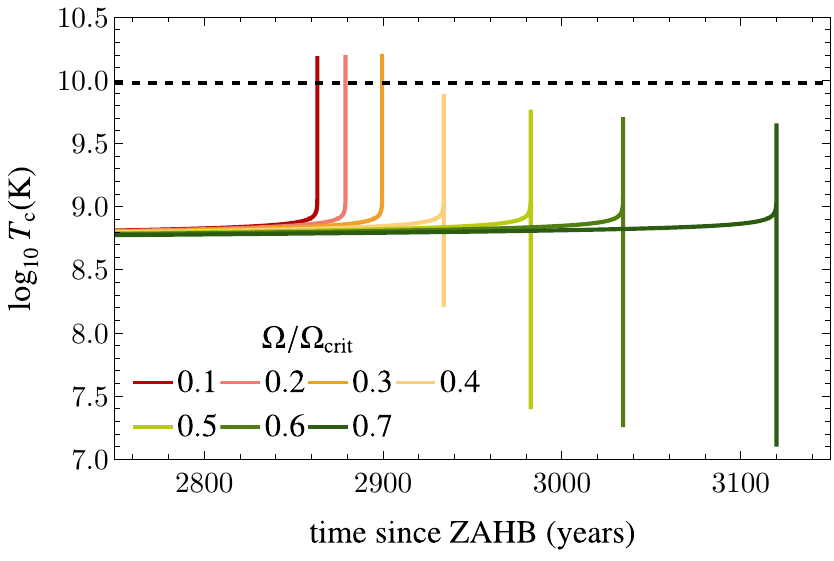}
    \caption{Central temperature $T_c$ versus {time since \vthree{core helium depletion}} for the simulations described in text, with $ M_{\rm in} = 160 \msun$ for the median $^{12}\mathrm{C}(\alpha,\gamma)^{16}\mathrm{O}$ rate.~Non-rotating models reach higher core densities and temperatures, crossing into the regime where photodisintegration reactions lead to gravitational collapse, here assumed to be $T_c = 9\times 10^9 \rm\, K$.~Rotating models, by contrast, terminate at lower $T_c$, thereby avoiding collapse and instead undergoing a PISN.}
    \label{fig:photodisintegration}
\end{figure}

\subsection{Implications for GW231123}

Our simulations imply that it is plausible to interpret GW231123’s primary component ($m_1\simeq137\msun$) as having formed via photodisintegration instability collapse, even with a rapidly rotating core in the absence of efficient angular momentum transport in the star.~A similar claim was previously made~\citep{Fishbach:2020qag} for the primary BH in GW190521, the former high‐mass record holder among GW events, although that interpretation depends strongly on the adopted prior distribution when analyzing the underlying GW data.

In our models, the ${}^{12}\mathrm{C}(\alpha,\gamma){}^{16}\mathrm{O}$ reaction rate needed to form GW231123's primary is both broad and realistic, ranging from the median to rates $3\sigma$ stronger.~The upper end is compatible with previous  hints  that higher rates are needed to explain GW observations e.g., the $35\msun$ peak in the BH mass function~\citep{Croon:2023kct}.~%

\vtwo{
While our analysis has focused on the primary component of GW231123, it is worth briefly commenting on the secondary BH. The inferred mass posterior for the secondary is compatible with formation below the gap --- the lower edge of the 90\% credibility interval is $m_2 \sim 50\,M_\odot$. A natural possibility is therefore that GW231123 straddles the mass gap. This interpretation is supported from a GW data analysis perspective: in a massive event like GW231123, the total mass $m_1+m_2$ is measured more accurately than either individual mass, which are consequently anti-correlated. If $m_1$ lies toward the upper end of its uncertainty range (placing it above the gap), then $m_2$ will lie toward the lower end of its range (and thus below the gap), \vthree{a mechanism previously suggested to explain GW190521 by \citet{Fishbach:2020qag}.}~As shown in this paper, this picture is further supported by the fact that rotation shifts the lower edge of the mass gap to higher masses. The secondary BH does not pose additional challenges to stellar-evolution models, although a dedicated evolutionary study of this object would certainly be worth pursuing.
} 

\section{Outlook}
GW231123 is a puzzling 
GW detection, with BHs that are among both the most massive and the most spinning detected so far.~The primary BH, in particular, can reach $m_1\sim 160 \msun$, and is thus a promising candidate for a BH ``beyond the gap,'' with a progenitor subject to the photodisintegration instability.~

We performed the first investigation of stars above the upper edge of the BHMG that includes both the effect of stellar rotation and variations in the critical  ${}^{12}\mathrm{C}(\alpha,\gamma){}^{16}\mathrm{O}$  nuclear-reaction rate.~{The main conclusion of this paper is that \vtwo{in the regime of weak angular momentum transport motivated by the spins of GW231123, it is possible that the  primary component of GW231123 formed from direct stellar collapse.}

\vtwo{
An important assumption of this analysis is that we do not include the ST dynamo or other mechanisms that enforce strong angular momentum coupling. Efficient coupling would drive the core toward uniform rotation and produce nearly non-spinning black holes (e.g. \citealt{Marchant:2020haw,Marchant:2023ncp}), which would be inconsistent with the high spin inferred for GW231123 under current interpretations. Our results should therefore be understood as an exploration of   
the BHMG in the presence of rapidly rotating stellar cores,
motivated by proposed scenarios in which massive stars can retain or regain substantial angular momentum.
}

{Previous studies have shown that} much of the information on population features often comes from a few highly informative events~\citep{2019MNRAS.484.4008G,Essick:2021vlx,Moore:2021xhn,Baxter:2021swn,Mancarella:2025uat}.~As was the case for GW190521 when analyzing data from the first three LIGO/Virgo/KAGRA observing runs~\citep{KAGRA:2021duu}, we anticipate that the exceptionally large masses and spins of GW231123 will play a major role in upcoming GW population constraints with O4 data.~\vtwo{Accordingly, our conclusions that the large masses of GW231123 can be produced in combination with large spins through stellar evolution in the weak-coupling regime provides a foundation for population‐level studies.}~{We anticipate that it will be important for future GW population analyses to allow the stellar component to accommodate systems like GW231123, rather than imposing a hard cutoff at the BHMG.~Fixing such a cutoff a priori could bias inferences toward exotic formation scenarios.}

\section*{Acknowledgements}

We thank Monica Colpi for discussions.~D.C.~is supported by STFC Grant No.~ST/T001011/1.~J.S.~is supported by NSF Grant No.~2207880.~D.G.~is supported by ERC Starting Grant No.~945155--GWmining, Cariplo Foundation Grant No.~2021-0555, MUR PRIN Grant No.~2022-Z9X4XS, Italian-French University (UIF/UFI) Grant No.~2025-C3-386, MUR Grant ``Progetto Dipartimenti di Eccellenza 2023-2027'' (BiCoQ), MSCA Fellowship No.~101064542--StochRewind, MSCA Fellowship No.~101149270--ProtoBH, MUR Young Researchers Grant No.~SOE2024-0000125, and the ICSC National Research Centre funded by NextGenerationEU.~J.S. thanks the IPPP for hospitality and acknowledges an IPPP DIVA fellowship to support the visit.~J.S. and D.G. thank the organizers of the International Congress of Basic Science (Beijing, 2025).~Our simulations were run on the University of Hawai\okina i's high-performance supercomputer KOA.~The technical support and advanced computing resources from University of Hawai\okina i Information Technology Services – Cyberinfrastructure, funded in part by the NSF MRI award No.~1920304, are gratefully acknowledged.~%

\section*{Data Availability}
Our code can be found in the paper's reproduction package at the following URL:~\href{https://zenodo.org/records/16898502}{https://zenodo.org/records/16898502}.
Software used includes: MESA version 15140, MESASDK version 20210401, GFORTRAN GCC version 9.2.0, 
GFORTRAN GCC version 9.2.0,
Jupyter Notebook version 6.4.12,
Python version 3.8.5,
Mathematica via Wolfram version 14.2.

\bibliographystyle{mnras_tex_edited}
\bibliography{refs} 

\appendix

\section{Sensitivity to collapse description}
\label{sensitivitytocollapse}
In Fig.~\ref{fig:collapsesens} we examine the dependence of the final BH mass and dimensionless spin parameter $\chi$ on the initial rotation rate $\Omega/\Omega_{\rm crit}$ for two different choices of the prompt‐collapse core mass, $M_{\rm cutoff}=3\,M_\odot$ and $5\,M_\odot$.~ The left panel shows that across the full range $0.10\le\Omega/\Omega_{\rm crit}\le0.40$, the resulting BH mass varies by at most $\Delta M\lesssim0.4\,M_\odot$, while the right panel demonstrates that the spin parameter changes by less than $\Delta \chi\lesssim0.02$.~ These small offsets confirm that our assumption of the mass of the inner core collapsing promptly has a negligible impact on the median‐rate outcomes.




\begin{figure}[t]
    \centering
    \includegraphics[width=\columnwidth]{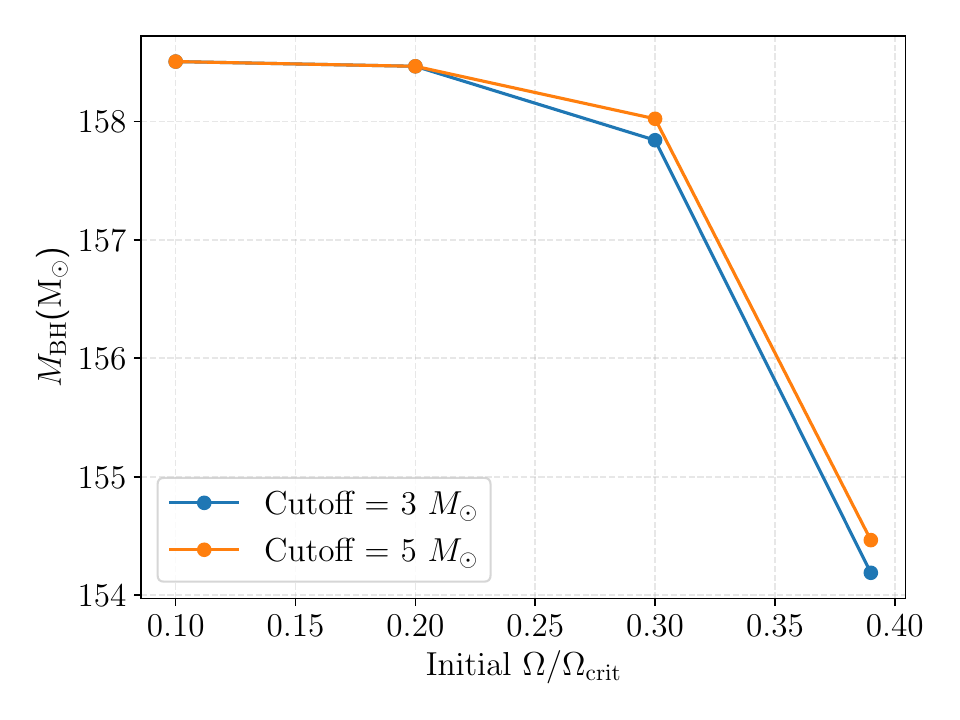}
    \includegraphics[width=\columnwidth]{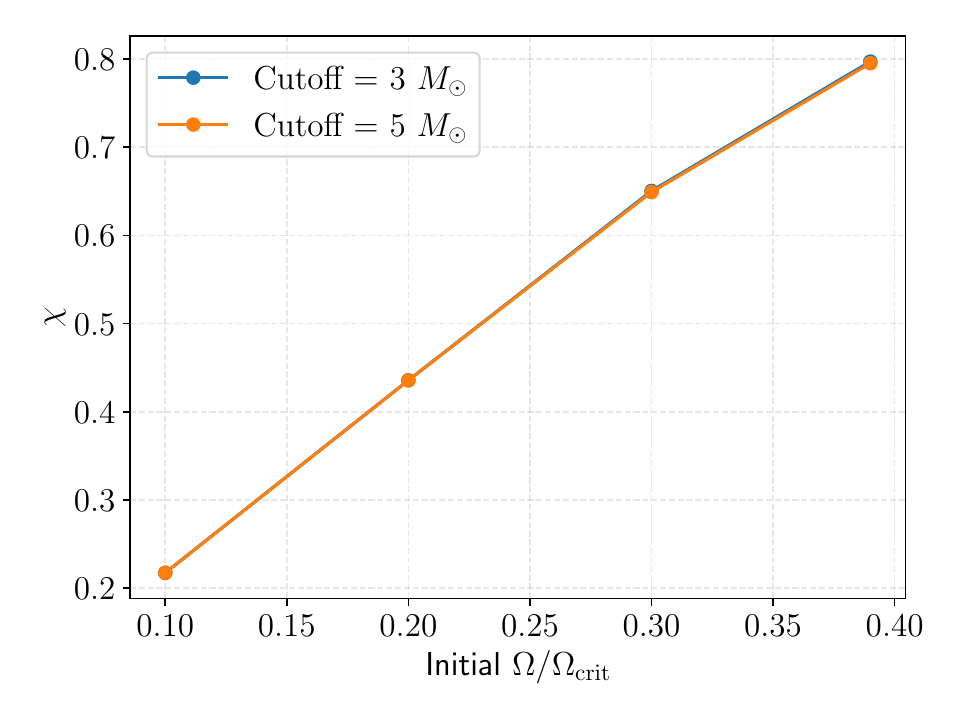}
    \caption{
    Final black‐hole mass (\textit{top}) and dimensionless spin parameter $\chi$ (\textit{bottom}) as functions of the initial rotation rate $\Omega/\Omega_{\rm crit}$ for two choices of the prompt‐collapse core mass threshold, $M_{\rm cutoff}=3\,M_\odot$ and $5\,M_\odot$.~
    }
    \label{fig:collapsesens}
\end{figure}
\end{document}